\documentclass[aps,showpacs,preprintnumbers,amsmath,amssymb,12pt,floatfix]{revtex4}

\usepackage{latexsym} 
\usepackage{amssymb}  
\usepackage{amsbsy}   
\usepackage{graphicx}       
\usepackage[dvips]{color}
\usepackage{bm}   

\newcommand{\p}{\partial}

\newcommand{\txt}{\textstyle}

\newcommand\eqn[1]{(\ref{#1})}      
\newcommand{\e}{{\rm e}}   

\def\tg{\tilde{\gamma}}

\newcommand{\be}{\begin{equation}}
\newcommand{\ee}{\end{equation}\noindent}
\newcommand{\bea}{\begin{eqnarray}}
\newcommand{\eea}{\end{eqnarray}\noindent}

\newcommand{\half} {{\txt \frac{1}{2}}}
\newcommand{\third}{{\txt \frac{1}{3}}}

\newcommand{\sixth}{{\txt \frac{1}{6}}}

\newcommand{\Det}{\mbox{Det}}

\newcommand{\nn}{\nonumber \\}
\newcommand{\gbar}{\bar\gamma}

\makeatletter 

\def\appendix{\par                              
    \setcounter{section}{0}                     
    \setcounter{subsection}{0}
    \renewcommand{\theequation}{\Alph{section}.\arabic{equation}}
    \renewcommand{\thesection}{Appendix \Alph{section}
                \setcounter{equation}{0}  } 
    \renewcommand{\thesubsection}{\Alph{section}.\arabic{subsection}}
}

\def\applabel#1{\@bsphack
  \protected@write\@auxout{}%
         {\string\newlabel{#1}{{\Alph{section}}{\thepage}}}%
  \@esphack}

\def\section{
\setcounter{equation}{0}        
\@startsection {section}{1}{\z@}{-3.5ex plus -1ex minus
 -.2ex}{2.3ex plus .2ex}{\large\bf}}
\renewcommand{\theequation}{\arabic{section}.\arabic{equation}}

\def\subsection{\@startsection{subsection}{2}{\z@}{-3.25ex plus -1ex minus
 -.2ex}{1.5ex plus .2ex}{\normalsize\bf}}

\def\subsubsection{\@startsection{subsubsection}{3}{\z@}{-3.25ex plus
 -1ex minus -.2ex}{1.5ex plus .2ex}{\normalsize}}

\makeatother   

\begin{document}

\title{Multidimensional Worldline Instantons}

\author{Gerald V.\ Dunne}
\author{Qing-hai Wang}
\affiliation{Department of Physics, University of Connecticut,
Storrs, CT 06269-3046, USA}

\date{August 2, 2006}

\begin{abstract}
We extend the worldline instanton technique to compute the vacuum pair production rate for spatially inhomogeneous electric background fields, with the spatial inhomogeneity being genuinely two or three dimensional, both for the magnitude and direction of the electric field. 
Other techniques, such as WKB, have not been applied to such higher dimensional problems. Our method exploits the instanton dominance of the worldline path integral expression for the effective action.
\end{abstract}

\begin{titlepage}
\maketitle
\renewcommand{\thepage}{}          

\end{titlepage}

\section{Introduction}
\label{intro}

A remarkable prediction of quantum electrodynamics (QED) is that in the presence of an electric field the polarization of the vacuum can lead to the production of electron-positron pairs from vacuum. This was initially predicted and estimated for approximately uniform fields \cite{he,schwinger}, but has not yet been observed experimentally due to the extreme scales required \cite{greiner,ringwald}. In achieving such extreme field strengths, the inhomogeneity of the field becomes important. Much work has been done to estimate the pair production rate for electric fields whose direction is fixed and whose magnitude varies in one dimension, either spatial \cite{nikishov,kimpage} or temporal \cite{nikishov2,brezin,popov}. These approaches are semiclassical, essentially WKB or its variants. More recently, a Monte Carlo worldline loop method has been developed and applied to the vacuum pair production problem \cite{giesklingmuller}. In principle, this Monte Carlo method can be applied to very general electric fields, but it has so far only been applied to the case of one-dimensional spatial inhomogeneities. The worldline instanton method, in which the Monte Carlo sum is effectively dominated by a single instanton loop, was introduced for constant fields in \cite{affleck}, and extended to inhomogeneous fields in \cite{wli1,wli2}. For one-dimensional inhomogeneities, the agreement between WKB methods, worldline instantons and the Monte Carlo results is excellent \cite{wli2}. In this paper, we apply the worldline instanton method to electric fields with multidimensional inhomogeneities. Specifically, we compute the pair production rate for a class of spatially inhomogeneous electric fields, in which both the magnitude and direction of the electric field vary in two or three dimensional space. It is not clear how to compute the pair production rate for such fields using WKB.

The vacuum pair production rate can be deduced from the imaginary part of the effective action $\Gamma$ \cite{schwinger}:
\be
P_{\rm production}=1-e^{-2\, {\rm  Im}\, \Gamma}\approx 2\, {\rm  Im}\, \Gamma\quad.
\label{decay}
\ee
The technical problem is thus to compute ${\rm  Im}\, \Gamma$ for a given background electric field. This is nontrivial as the answer is non-perturbative in the field. For example, for a constant electric field the leading weak field result (for scalar QED) is
\be
 \frac{{\rm  Im}\, \Gamma}{V_4} \sim \frac{e^2 E^2}{16\pi^3} \, e^{-\frac{m^2\pi}{e E}}\quad.
 \label{constant}
 \ee
 For inhomogeneous fields, the simplest approximation is the ``locally constant field'' (LCF) approximation, in which one replaces the constant $E$ in \eqn{constant} by $E(x)$, and integrates:
 \be
 {\rm  Im}\, \Gamma^{\rm LCF} \sim \frac{e^2}{16\pi^3} \int d^4 x\, \vec{E}^2(x)\, e^{-\frac{m^2\pi}{e | \vec{E}(x) |}}\quad.
 \label{lcf}
 \ee

In this paper we compute improved approximations to ${\rm Im}\,\Gamma$ using the worldline instanton method. In Sec.\ref{wli}, we review the worldline instanton method for computing the prefactor; in Sec.~\ref{2Dsym}, we present explicit computations of ${\rm Im}\,\Gamma$ for two-dimensional spatially inhomogeneous electric field backgrounds, and in Sec.~\ref{3Dsym}, we present explicit computations of ${\rm Im}\,\Gamma$ for three-dimensional spatially inhomogeneous electric field backgrounds. We conclude with some brief comments on possible extensions.

\section{Worldine Instanton Method}
\label{wli}

The worldline formalism for QED provides a nonperturbative approach to computing $\Gamma$, and in particular its imaginary part. The worldline formalism expresses the one-loop effective action 
for a scalar charged particle (of charge $e$ and mass $m$) in a gauge background $A_\mu$ as a quantum mechanical path integral \cite{feynman,halpern,bern,strassler,csreview}
\bea
\Gamma [A] =
\int_0^{\infty}\frac{dT}{T}\, \e^{-m^2T}\int d^4x^{(0)}
\!\!\!\!\!\!\! \int\limits_{x(T)=x(0)=x^{(0)}}\!\!\!\!\!\!\!\!\!\! {\mathcal D}x
\,\, {\rm exp}\left[-\int_0^Td\tau
\left(\frac{\dot x^2}{4} +i e A\cdot \dot x \right)\right] .
\label{eff}
\eea
Here the functional integral $\int {\mathcal D}x$ is over all closed Euclidean spacetime paths $x_\mu(\tau)$ that are periodic (with period $T$) in the proper-time parameter $\tau$. These closed paths are based at the marked point $x_\mu^{(0)}$, whose location is integrated over \cite{polyakov,gerry}. 
We use the path integral normalization conventions of \cite{csreview}. 
The effective action $\Gamma [A]$ is a functional of the classical background field $A_\mu(x)$, which is a given function of the space-time coordinates.

The worldline instanton method is based on the observation \cite{affleck,wli1,wli2}  that for certain classical background fields the quantum mechanical path integral in \eqn{eff} may be dominated by an instanton configuration, which is a closed path $x_\mu(\tau)$ satisfying the classical Euclidean equations of motion
\bea
\ddot{x}_\mu=2ie\, F_{\mu\nu}(x)\,\dot{x}_\nu \quad, \quad (\mu,\nu=1\dots4)\quad,
\label{euler}
\eea
where $F_{\mu\nu}=\partial_\mu A_\nu-\partial_\nu A_\mu$ is the background field strength.
{\it Worldline instantons} are periodic solutions to \eqn{euler}. For uniform \cite{affleck} and inhomogeneous \cite{wli1} background electric fields, these classical worldline instantons straightforwardly determine the nonperturbative exponential factor in ${\rm Im} \,\Gamma$.

However, this exponential factor is only part of the story. For inhomogeneous fields, the prefactor is also important. For example, with a spatially inhomogeneous electric field, there is a cutoff in the scale of the inhomogeneity, beyond which the  pair production rate vanishes. Interestingly,  this cutoff is reflected in the prefactor and not in the exponential factor \cite{giesklingmuller,wli2,kimpage}. Physically, this cutoff arises when the range of the electric field is so small that a virtual pair accelerated by the field does not acquire sufficient energy to become real particles.

To compute the prefactor one must compute the effect of the fluctuations about the classical worldline instanton path. The  strategy for computing the prefactor was explained in \cite{wli2}. The basic idea is that the quantum mechanical path integral in \eqn{eff} may be approximated as \cite{levit}
\bea
\int\limits_{x(T)=x(0)=x^{(0)}} \!\!\!\!\!\!\!\!{\mathcal D}x
\,\, {\rm exp}\left[-\int_0^Td\tau
\left(\frac{\dot x^2}{4} +i e A\cdot \dot x \right)\right]
\approx \frac{1}{(4\pi)^2}   \frac{e^{-S[x^{\rm cl}](T)}}{\sqrt{{\rm Det}\, \Lambda}}  \quad,
\label{semi}
\eea
where $\Lambda$ is the operator describing quadratic fluctuations about the worldline instanton path:
\bea
\Lambda_{\mu\nu}\equiv -\frac{1}{2}\,\delta_{\mu\nu}\, \frac{d^2}{d\tau^2}+i e\, F_{\mu\nu}(x)\, \frac{d}{d\tau}+ie\, \partial_\mu F_{\nu\rho}(x)\, \dot{x_\rho} \quad.
\label{lambda}
\eea
An important technical observation \cite{levit,kleinert} is that ${\rm Det}\, \Lambda$ can be expressed as the determinant of a {\it finite dimensional} matrix, whose entries consist of certain solutions to the Jacobi equations, 
\be
\Lambda_{\mu\nu}\, \eta_\nu=0 \quad,
\label{jacobi}
\ee
 evaluated at $\tau=T$. Specifically, we define the four independent solutions $\eta^{(\nu)}$ to \eqn{jacobi}, with initial value boundary conditions
\bea
\eta_\mu^{(\nu)}(0)=0\quad ; \quad \dot{\eta}_\mu^{(\nu)}(0)=\delta_{\mu\nu}  \quad .
\label{iv}
\eea
Then \cite{levit,kleinert}
\bea
{\rm Det}(\Lambda) = {\rm det}\left[ \eta_\mu^{(\nu)}(T)\right] \quad .
\label{gy}
\eea
This provides a simple and general numerical method for computing the semiclassical approximation in \eqn{semi}, for a given proper-time $T$. The next step \cite{wli2} is to evaluate the remaining $T$ integral in \eqn{eff} using steepest descents at a particular critical value $T_c$. This procedure was implemented explicitly in \cite{wli2} for inhomogeneous fields varying in one dimension, and the agreement with other computational methods was shown to be excellent. Here, in this paper, we extend this computational procedure explicitly to higher dimensional inhomogeneities.

\subsection{Numerical method for finding worldline instanton loops}

To be specific, we consider spatially inhomogeneous electric fields, $\vec{E}(\vec{x})$, for which we write the Euclidean gauge field as
\bea
A_4(\vec{x})=-i\frac{E}{k}\, f(k \vec{x})\quad .
\label{a4}
\eea
Here $k$ is a parameter related to the scale of the variation of the function $f$. By analogy with Keldysh's adiabaticity parameter \cite{keldysh} for time dependent ionization problems, we define the dimensionless inhomogeneity parameter
\bea
\tilde{\gamma}=\frac{mk}{e E} \quad .
\label{gammat}
\eea
More general fields are characterized by many scale parameters, but the form in \eqn{a4} is sufficient to illustrate our results. 

The first numerical step is to find the closed worldline instanton path, as a periodic solution to the classical Euclidean equations of motions
\bea
\ddot{x}_j &=& 2 e E \partial_j f\, \dot{x}_4 \quad,\quad(j=1,2,3)\quad,
\label{static1}
\eea
\bea
\ddot{x}_4 &=& -2 e E  \partial_j f \, \dot{x}_j \quad .
\label{static2}
\eea
To solve numerically  these equations, we need to start from a point on the solution curve. In principle, this step can be done by a tedious search. However, for the cases considered in this paper, all the solutions pass through the local maximum of the electric field. Without loss of generality, we choose this point to be the origin: $\vec{x}=(0, 0, 0)$. We exploit the gauge freedom to choose $f(\vec{0})=0$. 

Note that the $\ddot{x}_4$ equation \eqn{static2} can be integrated immediately to yield
\begin{equation}
\dot{x}_4=-\frac{2eE}{k}f(k\vec{x})\quad ,
\label{x4d}
\end{equation}
where we have chosen the integration constant to vanish in order to have a periodic solution. 
Now, recall that \eqn{euler} implies that $\dot{x}_\mu^2$ is constant on a classical solution:
\bea
\dot{x}_\mu^2=a^2 \quad .
\label{a}
\eea
Thus, the magnitude, $| \dot{\vec{x}}(0)|$,  of the ``3-velocity'' is equal to this constant $a$. But we do not know the initial direction of the 3-velocity. We use as shooting parameters the direction of the initial 3-velocity. For a one-dimensional problem where $f$ is a function of just one spatial coordinate, no shooting is required.
For a two-dimensional problem where $f$ is a function of two spatial coordinates, one shooting parameter is required. 
For a three-dimensional problem where $f$ is a function of all three spatial coordinates, two shooting parameters are required. 
Given a value of the parameter $a$ in \eqn{a}, we integrate the equations \eqn{static1} and \eqn{static2}, adjusting the initial direction of the 3-velocity until a closed loop is found. The total proper-time $T$ required to close the loop is clearly a function of the parameter $a$, so
\bea
T=T(a) \quad {\rm or} \quad a=a(T) \quad .
\label{aT}
\eea

\begin{figure}[ht]
\centering
\includegraphics[width=0.4\textwidth]{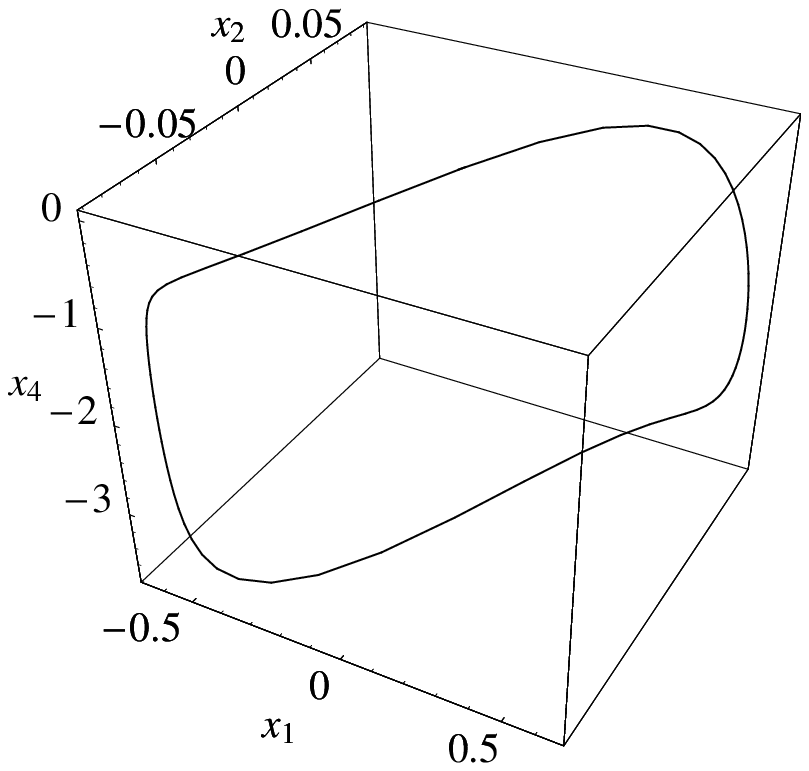}\\
\includegraphics[width=0.4\textwidth]{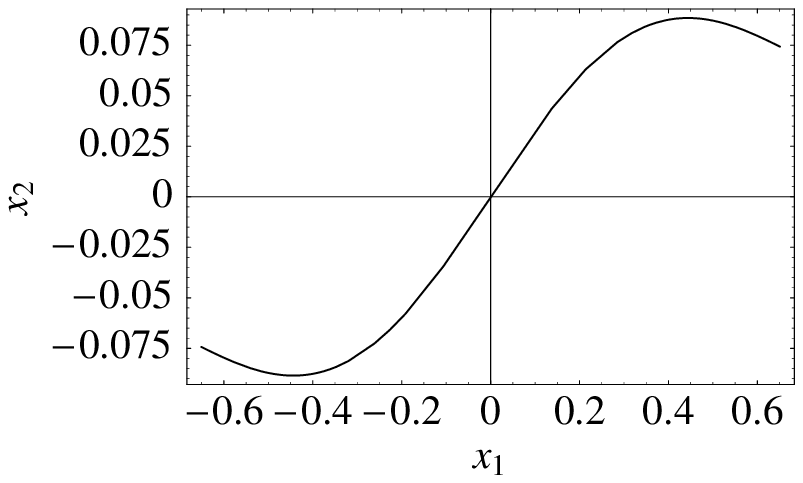}\\
\includegraphics[width=0.4\textwidth]{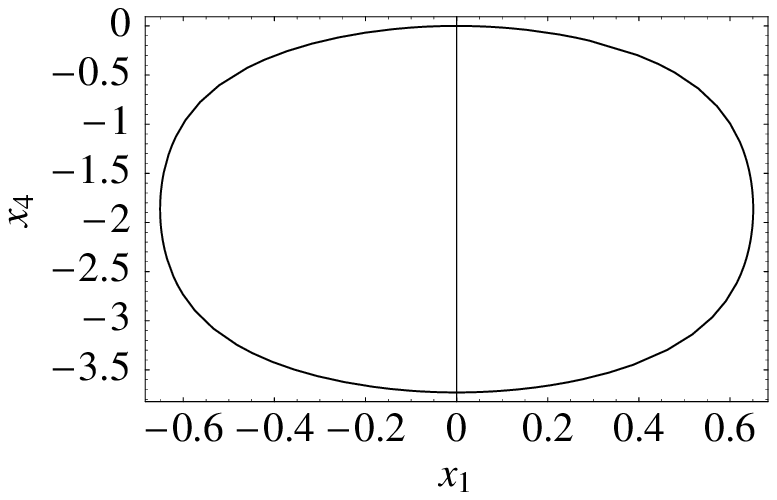}\\
\includegraphics[width=0.4\textwidth]{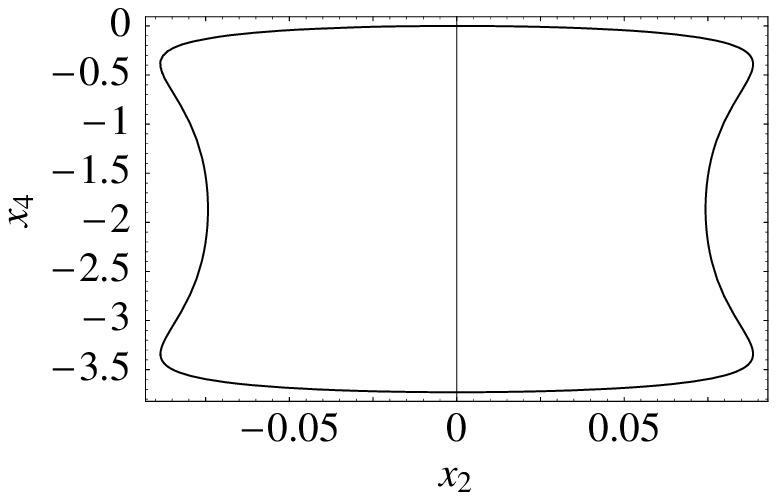}
\caption{Numerical worldline instanton loop for the case of $f(kx_1,kx_2)=\frac{\tanh(kx_1+kx_2)}{1+(kx_1)^2+10 (kx_2)^2}$.  
The first plot is the 3D parametrized plot of $\{x_1(\tau), x_2(\tau), x_4(\tau)\}$; the second plot is $x_1(\tau)$ vs.~$x_2(\tau)$; the third plot is $x_1(\tau)$ vs.~$x_4(\tau)$; and the last plot is $x_2(\tau)$ vs.~$x_4(\tau)$. The parameters used to generate these plots are $E=1$, $k=\sqrt{2}$, $a=0.593$, and $\dot{x}_1(0)=0.562\,408\,091\,043\,522\,23$.}
\label{fig:2D-loops}
\end{figure}

\begin{figure}[tb]
\centering
\includegraphics[width=0.3\textwidth]{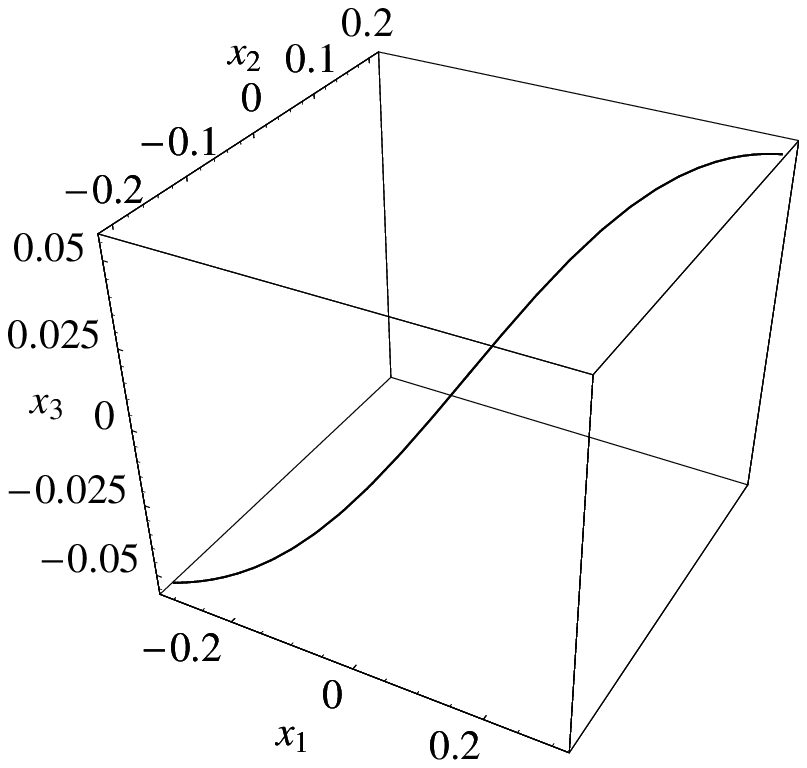}\\
\includegraphics[width=0.3\textwidth]{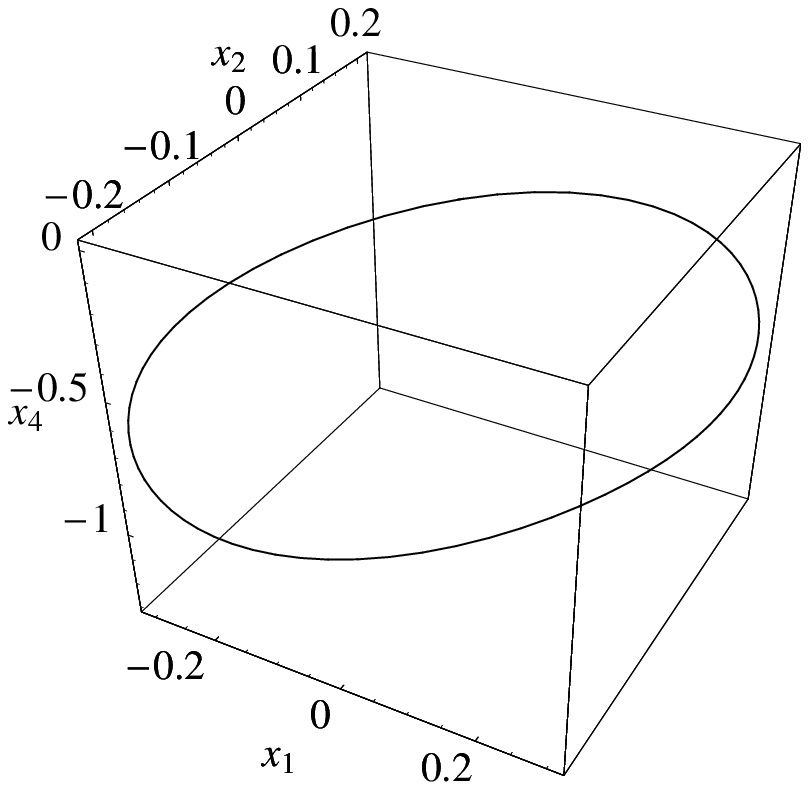}\\
\includegraphics[width=0.3\textwidth]{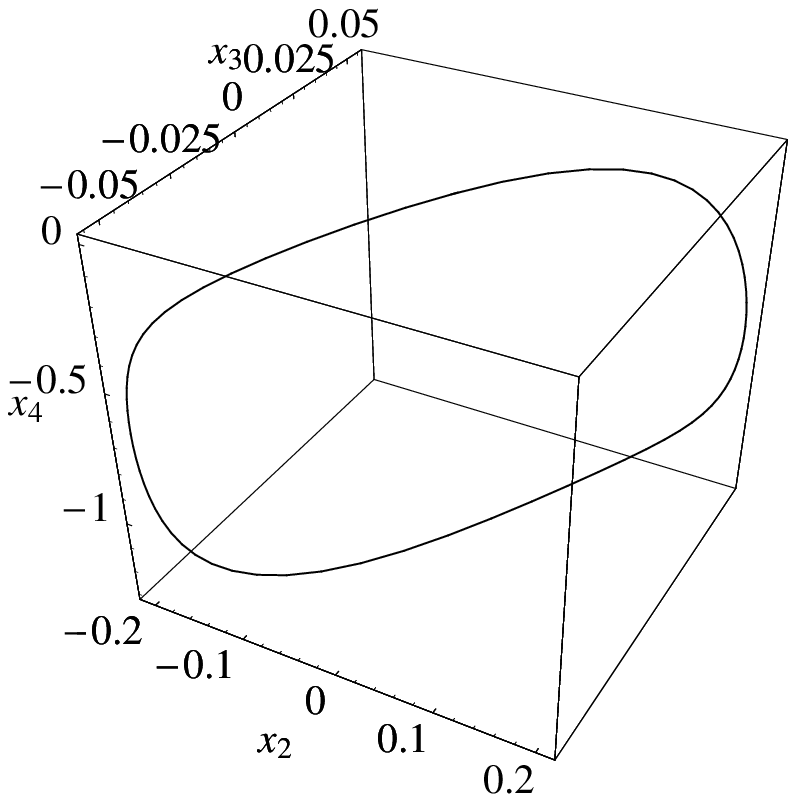}\\
\includegraphics[width=0.3\textwidth]{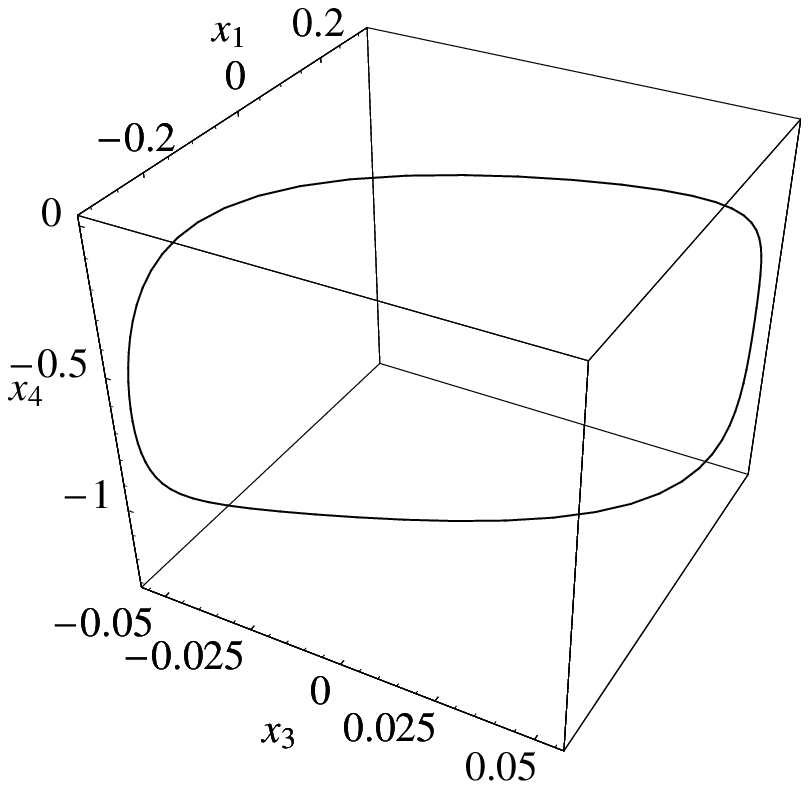}
\caption{Numerical worldline instanton loop for the case of $f(kx_1, kx_2, kx_3)=\frac{kx_1+kx_2+kx_3}{\left[1+(kx_1)^2+2 (kx_2)^2+10 (kx_3)^2\right]^2}$. 
The first plot is the 3D parametrized plot of $\{x_1(\tau), x_2(\tau), x_3(\tau)\}$; the second plot is $\{x_1(\tau), x_2(\tau), x_4(\tau)\}$; the third plot is $\{x_2(\tau), x_3(\tau), x_4(\tau)\}$; and the last plot is $\{x_3(\tau), x_1(\tau), x_4(\tau)\}$. The parameters used to generate these plots are $E=1$, $k=1$, $\dot{x}_1(0)=0.6$, $\dot{x}_2(0)=0.453\,000\,684$, and $\dot{x}_3(0)=0.201\,733\,127\,28$.}
\label{fig:3D-loops}
\end{figure}

For an electric field depending on two spatial coordinates, say $x_1$ and $x_2$, the loop has $\dot{x}_3=0$, and so depends in general on $x_1$, $x_2$ and $x_4$ only. An example of such a numerical worldline loop, for $f(kx_1,kx_2)=\frac{\tanh(kx_1+kx_2)}{1+(kx_1)^2+10 (kx_2)^2}$, is shown in Fig.~\ref{fig:2D-loops}. 
An example of such a numerical worldline loop for an electric field depending on all three spatial coordinates, with $f(kx_1, kx_2, kx_3)=\frac{kx_1+kx_2+kx_3}{\left[1+(kx_1)^2+2 (kx_2)^2+10 (kx_3)^2\right]^2}$, is shown in Fig.~\ref{fig:3D-loops}.
This loop depends on all four space-time coordinates, so we plot appropriate cross-sections.
Given these worldline loops, the pair production rate can be computed numerically as outlined above and in \cite{wli2}. In the following sections we illustrate this in some specific examples.

\section{Two-Dimensional Electric Fields}
\label{2Dsym}

\begin{figure}[tb]
\centering
\includegraphics[width=0.6\textwidth]{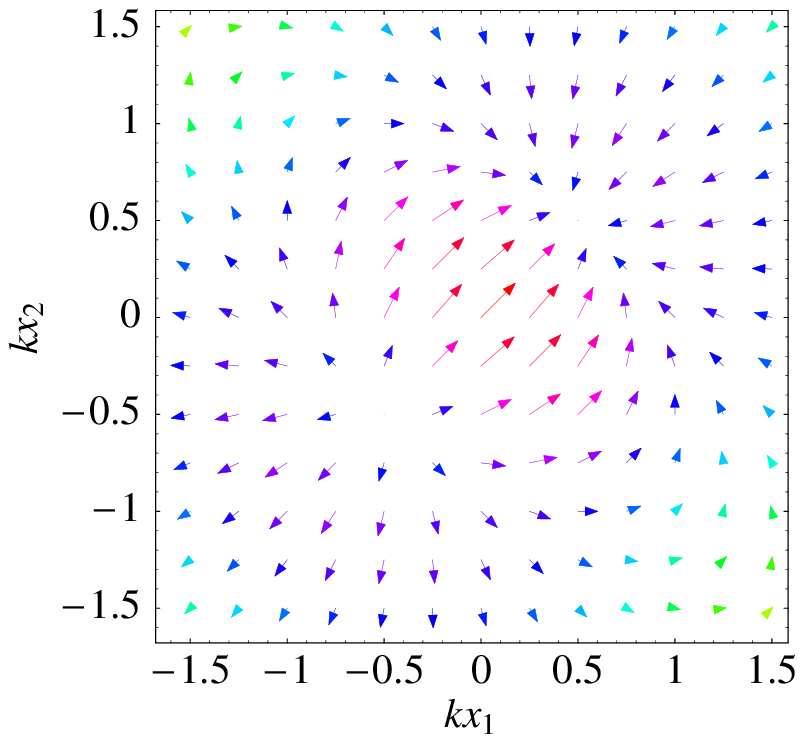}
\includegraphics[width=0.6\textwidth]{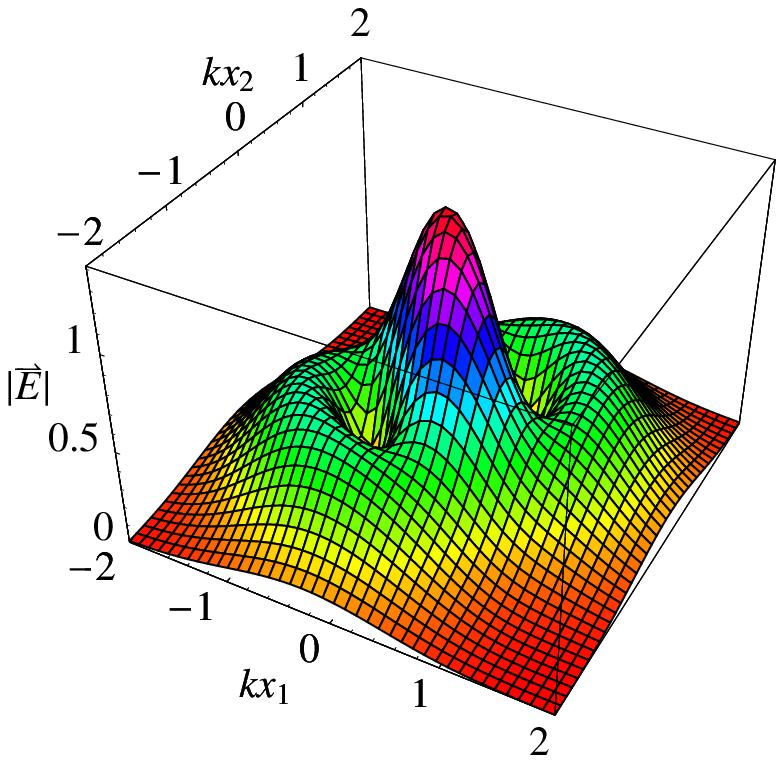}
\caption{The electric field $\vec{E}(x_1,x_2)$ for the case of $f(kx_1,kx_2)=(k x_1+k x_2)\e^{-(k x_1)^2-(k x_2)^2}$. The first plot is the vector plot of $\vec{E}$ as a function of $(kx_1, kx_2)$; the second plot is the 3D plot of the magnitude of $\vec{E}$ as a function of $(kx_1, kx_2)$. To generate these plots, we use the parameter $E=1$.}
\label{fig:2D-E}
\end{figure}

In this Section we illustrate our worldline instanton procedure in a
class of models where the electric field is static and only
depends on two spatial coordinates. For the 2D space-dependent fields of the form \eqn{a4}, the fluctuation operator \eqn{lambda} can be restricted to its components in the $(x_1, x_2, x_4)$ plane :
\begin{equation}
\Lambda = \left(\begin{array}{ccc}
-\frac{1}{2}\frac{d^2}{d^2\tau} + eEk f^{(2,0)}\dot{x}_4^{\rm cl} & eEk f^{(1,1)}\dot{x}_4^{\rm cl} & eEf^{(1,0)}\frac{d}{d\tau}\\
eEk f^{(1,1)}\dot{x}_4^{\rm cl} & -\frac{1}{2}\frac{d^2}{d^2\tau} + eEk f^{(0,2)}\dot{x}_4^{\rm cl} &
eE f^{(0,1)}\frac{d}{d\tau}\\
-eE\frac{d}{d\tau}f^{(1,0)} & - eE\frac{d}{d\tau}f^{(0,1)} & -\frac{1}{2}\frac{d^2}{d^2\tau}
\end{array}
\right)\quad,
\label{tfluc}
\end{equation}
where 
\be
f^{(m,n)}\equiv \frac{\p^m}{\p z_1^m} \frac{\p^n}{\p z_2^n} f(z_1,z_2)\quad.
\ee
To compute the fluctuation determinant we use the semiclassical quantum mechanical path integral result \eqn{gy}. Thus, we need to find solutions to the Jacobi equations $\Lambda\,\eta=0$ in \eqn{jacobi}, satisfying the initial value conditions \eqn{iv}. In general, this needs to be done numerically. However, there exists a class of fields for which the entire computation can be done semi-analytically, providing more insight.
Consider fields \eqn{a4} with the symmetry
\be
f(kx_1, kx_2)=f(k x_2,k x_1)\quad.
\label{2d-symmetric}
\ee
For example, the two-dimensional symmetric electric field example with $f(kx_1,kx_2)=(k x_1+k x_2)\e^{-(k x_1)^2-(k x_2)^2}$, is shown in Fig.~\ref{fig:2D-E}. 
As can be seen from the Figure, such an electric field is genuinely two dimensional in its inhomogeneity, both for its direction and its magnitude. Physically, we choose the form of $f$ so that the electric field is spatially localized. Interestingly, for such fields the worldline instanton loop, which is a closed loop in three dimensions [$x_1$, $x_2$, and $x_4$], is {\it planar}. This implies that much of the analysis can be reduced to the one dimensional case studied in \cite{wli2}. Note, however, that even though the classical loop is essentially 2D, the fluctuations extend non-trivially into the third direction. 

\subsection{Determinant of the Fluctuation Operator}

We write the solution of the Jacobi equation $\Lambda\phi=0$ as 
\begin{equation}
\phi=\left(
\begin{array}{c}
\phi_1\\
\phi_2\\
\phi_4
\end{array}
\right)\quad.
\end{equation}
Numerically, one simply integrates the Jacobi equations with the initial value conditions \eqn{iv}. However, we can partially construct these solutions analytically, as follows.
There are six linearly independent solutions, from which we need to construct the three independent solutions satisfying the initial conditions \eqn{iv}.
To find all six linearly independent solutions, we use two different {\it Ans\"atze}:
\begin{enumerate}
\item $\phi_1=\phi_2$. In this case, the Jacobi equation reduces to
\begin{eqnarray}
&&-\half\ddot{\phi}_1 + eEk\left[f^{(2,0)} + f^{(1,1)}\right]\dot{x}_4^{\rm cl}\phi_1 + eEf^{(1,0)}\dot{\phi}_4 = 0 \quad,\nn
&&\dot{\phi}_4 = -4eEf^{(1,0)}\phi_1 +v_4 \quad,
\end{eqnarray}
where $v_4$ is an integration constant. There are four independent solutions of this type. As in the 1D case considered in \cite{wli2}, we can write these four solutions as :
\begin{eqnarray}
\phi^{(1)}(\tau) &=& \left(\begin{array}{c}
0\\
0\\
1
\end{array}\right) \quad,\qquad \phi^{(2)}(\tau) = \left(\begin{array}{c}
\dot{x}_1^{\rm cl}(\tau)\\
\dot{x}_1^{\rm cl}(\tau)\\
\dot{x}_4^{\rm cl}(\tau)
\end{array}\right) \quad,\nn
\phi^{(3)}(\tau) &=& \left(\begin{array}{c}
\dot{x}_1^{\rm cl}(t)  \int_0^{\tau}dt\, \frac{1}{\left[\dot{x}_1^{\rm cl}(t)\right]^2}\\
\dot{x}_1^{\rm cl}(t)  \int_0^{\tau}dt\, \frac{1}{\left[\dot{x}_1^{\rm cl}(t)\right]^2}\\
\dot{x}_4^{\rm cl}(\tau)  \int_0^{\tau}dt\, \frac{1}{\left[\dot{x}_1^{\rm cl}(t)\right]^2} -  \int_0^{\tau}dt\, \frac{\dot{x}_4^{\rm cl}(t)}{\left[\dot{x}_1^{\rm cl}(t)\right]^2}
\end{array}\right) \quad,\nn
\phi^{(4)}(\tau) &=& \left(\begin{array}{c}
\dot{x}_1^{\rm cl}(\tau)  \int_0^{\tau}dt\, \frac{\dot{x}_4^{\rm cl}(t)}{\left[\dot{x}_1^{\rm cl}(t)\right]^2}\\
\dot{x}_1^{\rm cl}(\tau)  \int_0^{\tau}dt\, \frac{\dot{x}_4^{\rm cl}(t)}{\left[\dot{x}_1^{\rm cl}(t)\right]^2}\\
\dot{x}_4^{\rm cl}(\tau)  \int_0^{\tau}dt\, \frac{\dot{x}_4^{\rm cl}(t)}{\left[\dot{x}_1^{\rm cl}(t)\right]^2} -  a^2\int_0^{\tau}dt\, \frac{1}{\left[\dot{x}_1^{\rm cl}(t)\right]^2}
\end{array}\right) \quad .
\label{zeromodes1}
\end{eqnarray}
 
\item $\phi_1=-\phi_2$, and $\phi_4=0$. In this case, the Jacobi equation reduces to a single second order differential equation:
\begin{equation}
-\half\ddot{\phi}_1 + eEk\left[f^{(2,0)}-f^{(1,1)}\right]\dot{x}_4^{\rm cl}\phi_1  = 0 \quad.
\label{c6eqn}
\end{equation}
There are two independent solutions of this type. These are new to the two-dimensional case. Denote as $\psi^{(5)}$ and $\psi^{(6)}$ the two linearly independent solutions of \eqn{c6eqn} with the initial conditions:
\begin{eqnarray}
\psi^{(5)}(0)=1\quad, &\qquad& \dot{\psi}^{(5)}(0)=0\quad,\nn
\psi^{(6)}(0)=0\quad, &\qquad& \dot{\psi}^{(6)}(0)=1\quad.
\label{c6ic}
\end{eqnarray}
Then, we can write the last two solutions of the Jacobi solutions $\Lambda\phi=0$ as
\begin{equation}
\phi^{(5)}(\tau) = \left(\begin{array}{c}
\psi^{(5)}\\
-\psi^{(5)}\\
0
\end{array}\right) \quad,\qquad 
\phi^{(6)}(\tau) = \left(\begin{array}{c}
\psi^{(6)}\\
-\psi^{(6)}\\
0
\end{array}\right) \quad.
\label{zeromodes2}
\end{equation}

\end{enumerate}

Finally, given these six linearly independent solutions, $\phi^{(1)}, \dots, \phi^{(6)}$, we construct the linear combinations satisfying the initial conditions \eqn{iv} as
\begin{eqnarray}
\eta^{(1)}(\tau) &=& \half\dot{x}_1^{\rm cl}(0)\phi^{(3)}(\tau)+\half\phi^{(6)}(\tau)\quad, \nn
\eta^{(2)}(\tau) &=& \half\dot{x}_1^{\rm cl}(0)\phi^{(3)}(\tau)-\half\phi^{(6)}(\tau)\quad, \nn
\eta^{(4)}(\tau) &=& \half\dot{x}_4^{\rm cl}(0)\phi^{(3)}(\tau)-\half\phi^{(4)}(\tau) \quad .
\end{eqnarray}
A lengthy but straightforward computation shows that the fluctuation determinant \eqn{gy} takes the following simple form:
\bea
\Det (\Lambda) &=& T\det \left[\eta^{(1)}(T), \,\eta^{(2)}(T), \,\eta^{(4)} (T)\right]\nn
& =& \frac{T}{2}\frac{\left[\dot{x}_1^{\rm cl}(0)\right]^3}{ \dot{x}_1^{\rm cl}(T)} \psi^{(6)}(T)\left[a^2 I_1^2(T)-I_2^2(T)\right] \quad.
\label{det1}
\eea
Here, $I_1(T)$ and $I_2(T)$ are integrals defined in terms of the closed worldline instanton loop:
\begin{eqnarray}
I_1(\tau) &\equiv& \frac{\dot{x}_1^{\rm cl}(\tau)}{\dot{x}_1^{\rm cl}(0)}\, \int_0^\tau d t\,  \frac{1}{\left[\dot{x}_1^{\rm cl}(t)\right]^2} \quad,\nn
I_2(\tau) &\equiv&  \frac{\dot{x}_1^{\rm cl}(\tau)}{\dot{x}_1^{\rm cl}(0)}\, \int_0^\tau d t\,  \frac{\dot{x}_4^{\rm cl}(t)}{\left[\dot{x}_1^{\rm cl}(t)\right]^2} \quad .
\label{i1i2}
\end{eqnarray}
The overall factor of $T$ is from the free $x_3$ direction \cite{levit}.

The determinant \eqn{det1} can be simplified further using periodicity, which implies that $\dot{x}_1^{\rm cl}(0)=\dot{x}_1^{\rm cl}(T)$, and the vanishing of $I_2(T)$. Therefore, we find the simple expression for the fluctuation determinant:
\bea
\Det (\Lambda) = \frac{T}{2}\psi^{(6)}(T)\left[\frac{2eE}{k}\,\dot{x}_1^{\rm cl}(0)\, \gbar(T)\, I_1(T)\right]^2 \quad.
\label{det2}
\eea
Here we have defined
\bea
\bar{\gamma}(T)\equiv \frac{k}{2 e E}\, a(T) \quad .
\label{gamma-bar}
\eea

Now recall from \eqn{eff} and \eqn{semi} that we still need to evaluate the $4$-dimensional space-time integral over the fixed point on the closed loops:
\begin{eqnarray}
\int d^4 x^{(0)}&\equiv&\int dx_1(0)\, dx_2(0)\, dx_3(0)\, dx_4(0)\nn
&=& (L {\cal T}) \int dx_1(0)\, dx_2(0)\nn
&=& (L {\cal T}) \frac{N}{k} \int d\tau\, \dot{x}_1^{\rm cl}(0) \quad ,
\label{collective}
\end{eqnarray}
where $L$ is the $1$-space volume and ${\cal T}$ is the total time. $N$ is a normalization factor that depends on the orientation of the loop. The simplest way to evaluate $N$ is to compare the $k\to 0$ limit with the corresponding locally constant field approximation answer \eqn{lcf}, as explained below. Observe that the factor of $\dot{x}_1^{\rm cl}(0)$ appearing in \eqn{collective} cancels against the same factor in $\sqrt{\Det(\Lambda)}$ in \eqn{det2}, so that the spacetime integration effectively contributes a volume factor $ (L {\cal T})$, a factor of $N/k$, and a factor of $\frac{T}{2}$ from the $\tau$ integral. This last factor is just the collective coordinate contribution arising from invariance under shifts of the starting point on the loop, which gives rise to the second of the zero modes in \eqn{zeromodes1}, as discussed in the 1D case in \cite{wli2}.
Thus, collecting all the pieces, we see that the worldline instanton approximation \eqn{semi} to the quantum mechanical path integral leads to :
\begin{equation}
\Gamma^{\rm WLI}  \approx  L {\cal T} \, \frac{N}{(4\pi)^2}\,\frac{\sqrt{2}}{4eE}\,\int_0^{\infty}\frac{dT}{\sqrt{T}}\, \frac{\exp\left[-\left(S[x^{\rm cl}](T)+m^2 T\right)\right]}{\gbar(T)I_1(T)\sqrt{\psi^{(6)}(T)}} \quad ,
\label{tint}
\end{equation}
The main difference from the 1D case in \cite{wli2} is the appearance of the factor of $1/\sqrt{\psi^{(6)}(T)}$ and the power of $T$. The former is a measurement of the fluctuation in the second spatial direction; and the latter is because the remaining free dimension is now one instead of two.

\subsection{The $T$ integral}

As  in \cite{wli2}, for spatially inhomogeneous electric fields we use a rotated steepest descents method to evaluate the $T$ integral in the vicinity of a critical point $T_c$. The critical point is a stationary point of the exponent in \eqn{tint}:
\begin{equation}
\Delta(T) \equiv  S[x^{\rm cl}](T)+m^2 T\quad .
\label{exponent}
\end{equation}
This notation emphasizes the fact that the action $S[x^{\rm cl}]$, evaluated on the worldline instanton path $x^{\rm cl}(\tau)$, is a function of $T$. Near $T_c$, we have
\begin{equation}
\Delta(T)\approx \Delta(T_c)+\half\Delta''(T_c)(T-T_c)^2+\cdots
\end{equation}
As in the 1D cases, the critical point $T_c$ occurs when 
\begin{equation}
\gbar(T_c)=\tg \quad.
\label{tc}
\end{equation}
The rotation of the $T$ contour produces a factor of $i$, so our final expression is
\begin{equation}
{\rm Im}\, \Gamma^{\rm WLI} \sim (L{\cal T})\, \frac{N}{32\pi^{3/2}}\,\frac{1}{eE}\, \frac{\e^{-\Delta(T_c)}}{\tilde{\gamma}\,  I_1(T_c)\sqrt{T_c\,\psi^{(6)}(T_c)|\Delta''(T_c)|}} \quad .
\label{answer}
\end{equation}

\subsection{Locally Constant Field Approximation}

\begin{figure}[tb]
\centering
\includegraphics[width=0.7\textwidth]{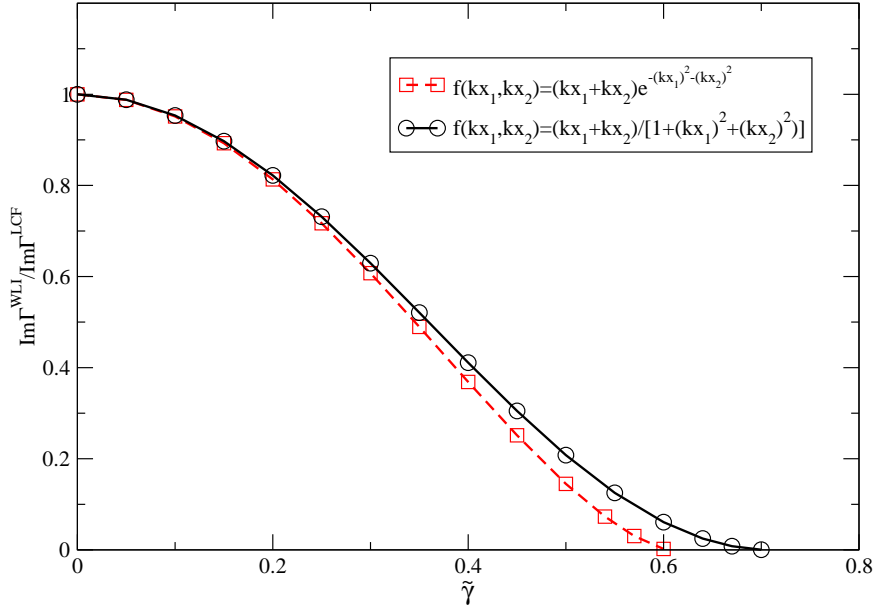}
\caption{${\rm Im}\,\Gamma^{\rm WLI}/{\rm Im}\,\Gamma^{\rm LCF}$ for the cases of $f(kx_1,kx_2)=\frac{kx_1+kx_2}{1+(kx_1)^2+(kx_2)^2}$, and $f(kx_1,kx_2)=(kx_1+kx_2)\e^{-(kx_1)^2-(kx_2)^2}$. The parameters used to generate this plot are $E=1$, $e=1$, and $m=1$.}
\label{fig:2D}
\end{figure}

As in \cite{giesklingmuller,wli2,kimpage}, we compare the worldline instanton result with the locally constant field approximation \eqn{lcf}. 
For example, consider the explicit example 
\begin{eqnarray}
f(kx_1,kx_2)=\frac{k x_1+kx_2}{1+(k x_1)^2+(k x_2)^2} \quad .
\label{fr2}
\end{eqnarray}
Expanding $|\vec{E}(x_1, x_2)|$ for small $k$, we find it is a quadratic function of a new pair of variables:
\begin{equation}
|\vec{E}(x_1,x_2)| \approx \sqrt{2}\,E \left(1-\frac{3}{2}k^2x_+^2-\frac{1}{2}k^2x_-^2+\cdots\right)\quad,
\end{equation}
where $x_+ \equiv x_1+x_2$, and $x_- \equiv x_1-x_2$.
Thus, the leading LCF approximation is
\begin{eqnarray}
{\rm Im}\,\Gamma^{\rm LCF} &\sim& (L{\cal T})\frac{(eE)^2}{32\pi^3}\int dx_+\,dx_- \exp\left[-\frac{m^2\pi}{\sqrt{2}\,eE}\left(1+\frac{3}{2}k^2x_+^2+\frac{1}{2}k^2x_-^2\right)\right]\quad,\nn
&=& (L{\cal T})\frac{(eE)^3}{4\sqrt{3}\,\pi^3(mk)^2}\e^{-\frac{m^2\pi}{\sqrt{2}\,eE}}\quad. 
\label{LCFanswer}
\end{eqnarray}
Comparing ${\rm Im}\,\Gamma^{\rm LCF}$ and the limiting value of ${\rm Im}\Gamma^{\rm WLI}$ as $k\to 0$, we find empirically that the normalization constant in \eqn{answer} for the case of \eqn{fr2} is
\begin{equation}
N=2^{5/4}\frac{\sqrt{eE}}{m} \quad .
\end{equation}
The same normalization factor arises for the exponentially localized field with $f(kx_1, kx_2)= (kx_1+kx_2)\e^{-(k x_1)^2-(k x_2)^2}$. The ratio of the 
worldline instanton answer \eqn{answer} to the LCF answer \eqn{LCFanswer} is plotted in Fig.~\ref{fig:2D} for two different forms of the 2D electric field. In each case, there is a smooth dependence of the ratio from $1$ to $0$ as the inhomogeneity scale $\tilde{\gamma}$ [{\it i.e.} $k$] increases, which corresponds to the range of the field decreasing. Physically, the vanishing of the imaginary part at a critical inhomogeneity scale is because for very short range electric fields virtual dipole pairs do not acquire sufficient energy when accelerated in the field in order to become real particles. This has been observed previously for one dimensional spatial inhomogeneities \cite{giesklingmuller,wli2,kimpage}, and here we see it also in the higher dimensional case.

\section{Three-Dimensional Electric Fields}
\label{3Dsym}

\begin{figure}[tb]
\centering
\includegraphics[width=0.7\textwidth]{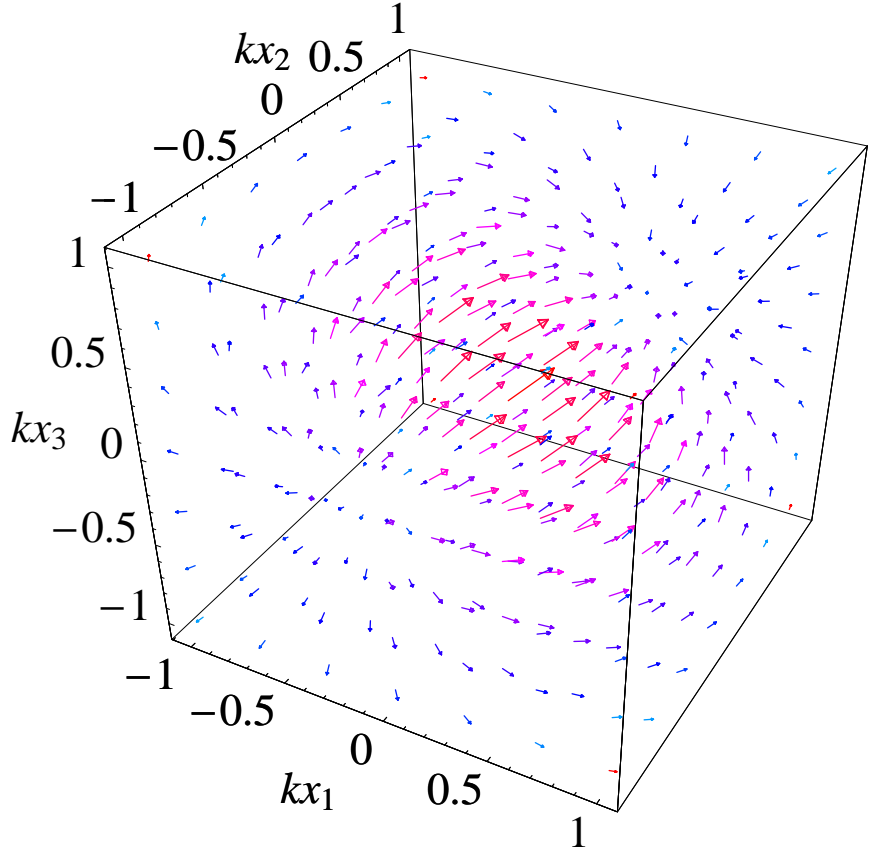}
\caption{Vector plot of the electric field $\vec{E}(x_1,x_2,x_3)$ for the case of $f(kx_1,kx_2,kx_3)=\frac{kx_1+kx_2+kx_3}{\left[1+(kx_1)^2+(kx_2)^2+(kx_3)^2\right]^{3/2}}$.}
\label{fig:3D-E}
\end{figure}

For three dimensions the computation is very similar, with just a few changes. First, finding the worldline instanton loop requires a two parameter shooting procedure. Second, the fluctuation operator $\Lambda$ in \eqn{lambda} is now a $4\times 4$ matrix differential operator. The computation can be done numerically, but as in the 2D case we can gain more insight by specializing to a class of fields for which ${\rm Im}\,\Gamma$ can be computed semi-analytically. 
Consider totally symmetric fields \eqn{a4} with
\be
f(kx_1, kx_2, kx_3)=f(kx_2, kx_1, kx_3)=f(kx_1, kx_3, kx_2)=f(kx_3, kx_2, kx_1)\quad.
\label{3d-symmetric}
\ee
The electric field for an example with $f(kx_1, kx_2, kx_3)=\frac{kx_1+kx_2+kx_3}{\left[1+(kx_1)^2+(kx_2)^2+(kx_3)^2\right]^{3/2}}$ is shown in Fig.~\ref{fig:3D-E}.
Note that it is localized, and that both the magnitude and direction of the field vary in three dimensions.

The calculation is similar to the two-dimensional symmetric case, except we now need eight [rather than six] independent solutions to the Jacobi equation, $\Lambda \phi=0$, in order to construct the four [rather than three] independent solutions $\eta^{(\nu)}$ satisfying the initial value conditions \eqn{iv}. In order to find the extra two solutions we extend the {\it Ans\"atze} for $\phi=(\phi_1, \phi_2, \phi_3, \phi_4)$ to:
\begin{enumerate}
\item Suppose $\phi_1=\phi_2=\phi_3$. Then $\Lambda \phi=0$ reduces to:
\begin{eqnarray}
&&-\half\ddot{\phi}_1 + eEk\left[f^{(2,0,0)}+2f^{(1,1,0)}\right]\dot{x}_4^{\rm cl}\phi_1 + eEk f^{(1,0,0)}\dot{\phi}_4 = 0 \quad,\nn
&&\dot{\phi}_4 = -6eE f^{(1,0,0)}\phi_1 +v_4 \quad.
\end{eqnarray}
There are four solutions of this type. These solutions of the Jacobi equation are similar to those in the one- and two-dimensional cases.
\item Suppose $\phi_1=-\phi_2$ and $\phi_3=\phi_4=0$. In this case, the Jacobi equation reduces to a single second order differential equation, which is very similar to \eqn{c6eqn}:
\begin{equation}
-\half\ddot{\phi}_1 + eEk\left[f^{(2,0,0)}-f^{(1,1,0)}\right]\dot{x}_4^{\rm cl}\phi_1  = 0 \quad.
\label{c6eqn3D}
\end{equation}
There are two solutions of this type. Define $\psi^{(5)}$ and $\psi^{(6)}$ to be the solutions to \eqn{c6eqn3D} with the initial conditions in \eqn{c6ic}. Then two further solutions can be written as
\begin{eqnarray}
\phi^{(5)}(\tau) = \left(\begin{array}{c}
\psi^{(5)}\\
-\psi^{(5)}\\
0\\
0
\end{array}\right) \quad, &\qquad& 
\phi^{(6)}(\tau) = \left(\begin{array}{c}
\psi^{(6)}\\
-\psi^{(6)}\\
0\\
0
\end{array}\right) \quad.
\end{eqnarray}
\item Suppose $\phi_1=\phi_2=-\half\phi_3$, and $\phi_4=0$. In this case, the Jacobi equation reduces to exactly the same equation as in \eqn{c6eqn3D}. There are two further solutions of this type. They can also be written in terms of $\psi^{(5)}$ and $\psi^{(6)}$:
\begin{eqnarray}
\phi^{(7)}(\tau) = \left(\begin{array}{c}
\psi^{(5)}\\
\psi^{(5)}\\
-2\psi^{(5)}\\
0
\end{array}\right) \quad, &\qquad& 
\phi^{(8)}(\tau) = \left(\begin{array}{c}
\psi^{(6)}\\
\psi^{(6)}\\
-2\psi^{(6)}\\
0
\end{array}\right) \quad.
\end{eqnarray}
\end{enumerate}

Given these eight linearly independent solutions to $\Lambda \phi=0$, the linear combinations satisfying the initial conditions \eqn{iv} are
\begin{eqnarray}
\eta^{(1)}(\tau) &=& \third\dot{x}_1^{\rm cl}(0)\phi^{(3)}(\tau)+\half\phi^{(6)}(\tau)+\sixth\phi^{(8)}\quad, \nn
\eta^{(2)}(\tau) &=& \third\dot{x}_1^{\rm cl}(0)\phi^{(3)}(\tau)-\half\phi^{(6)}(\tau)+\sixth\phi^{(8)}\quad, \nn
\eta^{(3)}(\tau) &=& \third\dot{x}_1^{\rm cl}(0)\phi^{(3)}(\tau)-\third\phi^{(8)}(\tau) \quad, \nn 
\eta^{(4)}(\tau) &=& \third\dot{x}_4^{\rm cl}(0)\phi^{(3)}(\tau) - \third\phi^{(4)}(\tau) \quad .
\end{eqnarray}
A lengthy but straightforward computation shows that the fluctuation determinant \eqn{gy} takes the simple form:
\bea
\Det (\Lambda) &= & \det \left[\eta^{(1)}(T), \,\eta^{(2)}(T), \,\eta^{(3)}(T), \,\eta^{(4)} (T)\right]\nn
& =& \frac{1}{3}\frac{\left[\dot{x}_1^{\rm cl}(0)\right]^3}{ \dot{x}_1^{\rm cl}(T)} \left[\psi^{(6)}(T)\right]^2\left[a^2 I_1^2(T)-I_2^2(T)\right] \quad,
\eea
where $I_1$ and $I_2$ are defined in \eqn{i1i2}. Notice the remarkable similarity to the 1D and 2D results for ${\rm Det} (\Lambda)$, in \cite{wli2} and \eqn{det2}, respectively.

Proceeding as in the 2D case, we find the following worldline instanton expression for the imaginary part of the effective action:
\begin{equation}
{\rm Im}\, \Gamma^{\rm WLI} \sim {\cal T}\, \sqrt{\frac{3}{2}}\,\frac{N}{32\pi^{3/2}k}\,\frac{1}{eE} \frac{\e^{-\Delta(T_c)}}{\tilde{\gamma}\,  I_1(T_c) \psi^{(6)}(T_c)\,\sqrt{|\Delta''(T_c)|}} \quad .
\label{answer-3d}
\end{equation}
Here $N$ is a normalization constant coming from the spatial integrals, which depends on the form of the planar worldline instanton loop. As before, the simplest way to evaluate $N$ is by comparison of the $k\to 0$ limit of ${\rm Im}\, \Gamma^{\rm WLI}$ with the locally constant field approximation expression in \eqn{lcf}. For example, for the case of  $f(kx_1,kx_2,kx_3)=(kx_1+kx_2+kx_3)\e^{-(kx_1)^2-(kx_2)^2-(kx_3)^2}$ we find 
\begin{equation}
{\rm Im}\,\Gamma^{\rm LCF} \sim {\cal T}\frac{3^{9/4}}{16\sqrt{7}}\, \frac{(eE)^{7/2}}{\pi^3(mk)^3} \e^{-\frac{m^2\pi}{\sqrt{3}eE}}\quad,
\label{LCFanswer-3dexp}
\end{equation}
which leads to $N=\frac{9}{\sqrt{7}}\, \frac{eE}{m^2}$. And for the case of $f(kx_1,kx_2,kx_3)=\frac{kx_1+kx_2+kx_3}{\left[1+(kx_1)^2+(kx_2)^2+(kx_3)^2\right]^{3/2}}$ we find 
\begin{equation}
{\rm Im}\,\Gamma^{\rm LCF} \sim {\cal T}\frac{3^{3/4}}{4\sqrt{14}}\, \frac{(eE)^{7/2}}{\pi^3(mk)^3} \e^{-\frac{m^2\pi}{\sqrt{3}eE}}\quad,
\label{LCFanswer-3d}
\end{equation}
and $N=\frac{6}{\sqrt{7}}\,\frac{eE}{m^2}$. 

\begin{figure}[tb]
\centering
\includegraphics[width=0.7\textwidth]{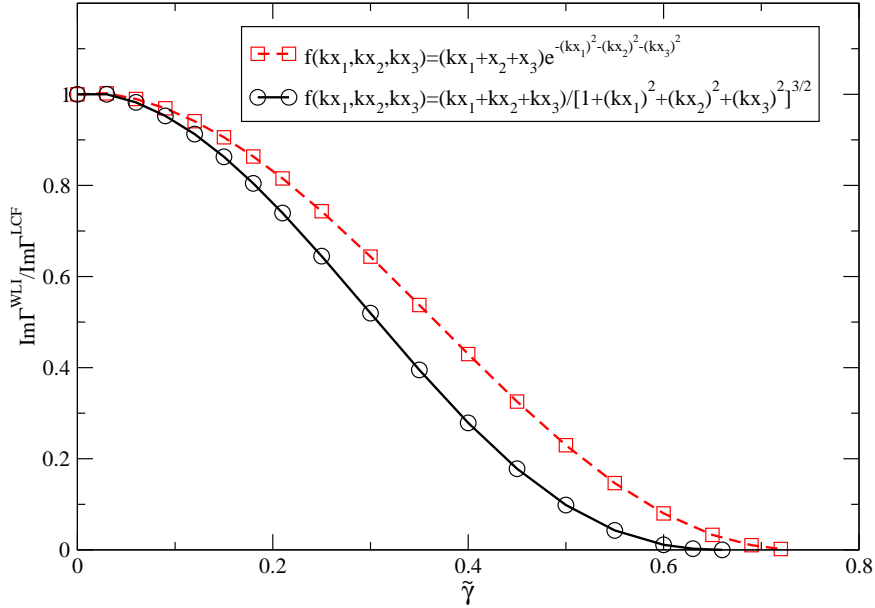}
\caption{${\rm Im}\,\Gamma^{\rm WLI}/{\rm Im}\,\Gamma^{\rm LCF}$ for the cases of $f(kx_1,kx_2,kx_3)=\frac{kx_1+kx_2+kx_3}{\left[1+(kx_1)^2+(kx_2)^2+(kx_3)^2\right]^{3/2}}$, and $f(kx_1,kx_2,kx_3)=(kx_1+kx_2+kx_3)\e^{-(kx_1)^2-(kx_2)^2-(kx_3)^2}$. The parameters used to generate this plot are $E=1$, $e=1$, and $m=1$.}
\label{fig:3D}
\end{figure}

In Fig.~\ref{fig:3D} we plot as a function of the inhomogeneity scale $\tg$, the ratio of the worldline instanton answer \eqn{answer-3d} to the LCF approximation answer \eqn{LCFanswer-3dexp} or \eqn{LCFanswer-3d}, for these fields. Note once again that the pair production rate vanishes when the field becomes too closely localized.

\section{Conclusions}
\label{conclusions}

In this paper we have extended the worldline instanton method of \cite{wli1,wli2} to the multi-dimensional case. Specifically, we have computed the imaginary part of the effective action in scalar QED for background electric fields that are inhomogeneous, both in their direction and magnitude, in two and three spatial dimensions. As in the one dimensional cases, the results exhibit the vanishing of the imaginary part when the inhomogeneity scale becomes too small, consistent with the physical picture of pair production as the acceleration of virtual dipole pairs in the electric field. In these multi-dimensional cases, there are currently no other computations with which to compare, except the crude locally constant field approximation, which does not capture the physics of the vanishing of ${\rm Im}\,\Gamma$ at short inhomogeneity scales. 
It would, however, be very interesting to compare with the numerical Monte Carlo worldline loop approach of Gies {\it et al} \cite{giesklingmuller}, to see if the instanton dominance of the worldline instanton method can be combined with the versatility of the Monte Carlo approach. Other interesting open questions include: (i) more general Abelian fields representing realistic laser fields \cite{ringwald}, with both spatial and temporal inhomogeneities; (ii) incorporating kinetic effects \cite{kluger,roberts} and backreaction effects \cite{cooper}; (iii) the extension to non-Abelian theories \cite{nussinov}, such as are relevant for heavy ion physics \cite{dima,raju}, and where the worldline instanton equations are the Wong equations \cite{wong}. In these more general cases, finding the worldline instanton closed loops will most likely require going beyond simple shooting techniques, to use action minimization algorithms.

\vskip 1cm

{\bf Acknowledgements:} We thank H.~Gies and C.~Schubert for discussions and comments, and we thank the DOE for support through the grant DE-FG02-92ER40716, and the NSF for support through the US-Mexico Collaborative Research Grant 0122615.

\end{document}